\documentclass{aa} 
\usepackage{natbib}
\usepackage{url}
\usepackage{graphicx}
\usepackage{subfig}
\usepackage{float}
\usepackage{txfonts}
\begin{document}

   \title{Spitzer observations of a circumstellar nebula around the candidate Luminous Blue Variable MWC 930}
 \titlerunning{A circumstellar nebula around MWC 930}
 \authorrunning{L. Cerrigone et al.}
   \author{L. Cerrigone
          \inst{1}
          \and
        G. Umana\inst{2} \and  C. S. Buemi\inst{2} \and  J. L. Hora\inst{3} \and  
          C. Trigilio\inst{2} \and  P. Leto\inst{2} \and A. Hart\inst{3} 
          }

   \institute{Centro de Astrobiolog{\'\i}a, INTA-CSIC
              ctra de Ajalvir km 4 Torrej\'on de Ardoz, E-28850 Spain\\
              \email{cerrigone@cab.inta-csic.es}
         \and
             INAF-Catania Astrophysical Observatory, via S. Sofia 78, I-95123 Catania, Italy\\
          \and
         Harvard-Smithsonian Center for Astrophysics, 38 Garden St., Cambridge, MA, USA, 02138\\
             }

   \date{}

 
  \abstract
   {MWC 930 is a star just $\sim$2$^\circ$ above the Galactic plane whose nature is not clear and that has not been studied in detail so far. While a post-Asymptotic Giant Branch (AGB) classification was proposed in the past, studies of its optical spectrum and photometry pointed toward strong variability, therefore the object was reclassified as a Luminous Blue Variable (LBV) candidate.}
   {LBVs typically undergo phases of strong mass loss in the form of eruptions that can create shells of ejecta around the star. Our goal is to search for the presence of such a circumstellar nebula in MWC 930 and investigate its properties.}
   {To do so, we make use of space-based infrared data from our Spitzer campaign performed with the InfraRed Array Camera (IRAC) and the InfraRed Spectrograph (IRS) as well as data from optical and infrared (IR) surveys.}
   {In our Spitzer images, we clearly detect an extended shell around  MWC 930 at wavelengths longer than 5 $\mu$m. The mid-infrared spectrum is dominated by the central star and mostly shows forbidden lines of [FeII], with an underlying continuum that decreases with wavelength up to $\sim$15 $\mu$m and then inverts its slope, displaying a second peak around 60 $\mu$m, evidence for cold dust grains formed in a past eruption. By modeling the SED, we identify two central components, besides the star and the outer shell. These extra sources of radiation are interpreted as material close to the central star, maybe due to a recent ejection. Features of C-bearing molecules or grains are not detected.}
   {}


   \maketitle
%

\section{Introduction}
Luminous Blue Variables (LBVs) are rare massive ($M_{ZAMS}\gtrsim 25\, M_\odot$) stars with high luminosities ($L \ge 10^{5.4} L_\odot$) subject to va\-ria\-bi\-li\-ty. Two groups of LBVs are identified. The first one includes those stars that display spectral and photometric variability  similar to that of S Dor, the classical LBV prototype. These sources are subject to episodic outbursts with variations of 1--2 magnitudes at approximately constant bo\-lo\-metric lu\-mi\-no\-si\-ty over a time of years or decades. The second group includes stars characterized by giant eruptions, when the visual brightness of the star grows of an order of magnitude and the bolometric luminosity is also increased \citep{humphreys}. The latter group contains only P Cyg and $\eta$ Car, in the Galaxy.

The origin of the instability of these stars is not clear, but it is commonly accepted that it must be related to them being close to the Eddington limit for stability against radiation pressure \citep{nota}. The limit is expressed as the dimensionless $\Gamma$ parameter, defined as the ratio of radiative to gravitational acceleration, which is proportional to the ratio of stellar luminosity to mass  ($\Gamma_{Edd} \propto L/M$).  In fact, the less luminous members of the group can be as close to the Eddington limit as the more luminous members, if they have undergone heavy mass loss during a previous red supergiant phase. 

From an evolutionary point of view, LBVs are considered as transition objects between O stars and Wolf-Rayet (WR) stars, which will eventually explode as Supernovae (SN), although this evolutionary path has been recently challenged by observations that indicate how (some?) LBVs may directly go into the SN phase \citep{vink}.

Our knowledge of the LBV phenomenon is hampered by the paucity of stars that display it. So far, only 14 such stars are confirmed LBVs in our Galaxy \citep{vink}. To be confirmed as an LBV, a star must at least show both the spectral-type and photometric variability observed in S Dor-type sources, keeping in mind that this phenomenon may be intermittent in a star's life. The presence of a circumstellar nebula around a supergiant is evidence for past high mass loss by the star and can be taken as an indication that the star is an LBV candidate.

\subsection{MWC 930}
Optical spectra of MWC 930 display Balmer lines in emission as well as permitted and forbidden emission lines of $[$FeII$]$ \citep{partha}. Using low-resolution spectra, \citet{gauba} detected H$\alpha$ in emission and a weak circumstellar nebulosity of less than 2$''$. They concluded that the target may be a hot post-AGB star at a distance of about 200 pc. 

\citet{miros} performed long-term photometric monitoring of the source and acquired high-resolution optical spectra. In their data, they found strong photometric va\-ria\-bi\-li\-ty (inappropriate of a post-AGB star) and several spectral lines with P Cyg profiles, pointing to a terminal wind velocity of $\sim$140 km~s$^{-1}$. They also estimated a lower limit to the mass-loss rate of $1.5\times 10^{-6} M_\odot yr^{-1}$, which also does not match the spectral characteristics of a star in the post-AGB evolutionary stage. Their spectral analysis indicates  a central star temperature of $22\,000\pm5\,000$ K and, with their estimate of the distance to MWC 930 of 3.5 kpc, a luminosity of $10^{5.5\pm0.2} L_\odot$. These values rule out a post-AGB classification and rather make the star a candidate LBV.

Subsequent studies  confirmed the spectral classification as a B5--B9 star with emission lines showing P Cyg profiles  \citep{carmona} as well as Brackett emission lines in the near-IR, without detectable CO bands \citep{raman}.

\section{Observations and results}
We observed MWC 930  with the InfraRed Array Camera \citep[IRAC;][]{irac} and the InfraRed Spectrograph \citep[IRS;][]{irs} on-board the Spitzer Space Telescope \citep{spitzer} within program 50116 (PI: G. Fazio). The IRAC images were obtained on 2009 May 01 (AOR 25444608)  and the IRS spectra two days later (AOR 25444352). For both instruments, the pointing coordinates were 18:26:25.24 in right ascension and -07:13:17.7 in declination (J2000).

\subsection{Spitzer IRAC observations}
The infrared imaging was performed  at 3.6, 4.5, 5.8, and 8.0 $\mu$m in High Dynamic Range mode. For each target we obtained eight dithered frames, for a total on-source time of about 96 s per IRAC channel. Corrected Basic Calibrated Data (cBCDs)  were retrieved from the Spitzer archive (pipeline version S18.25.0) and then coadded with a sampling of 0.3$''$ per pixel using MOPEX \citep{mopex}. These are BCDs where artefact mitigation software has been applied. Photometry on the central source was performed by point-spread function (PSF) subtraction in the final mosaics with IRACproc (Schuster et al, 2006). 

Since the central star is quite bright, we extracted our photometry using PSFs that were specifically developed to perform photometry even on heavily saturated objects. The PSFs had been obtained from observations of bright stars and then directly normalized to the IRAC observations of Vega \citep{marengo06, marengo07}. Therefore, by fitting the low-level features (diffraction spikes and wings) of the saturated PSFs of the observed source, magnitudes can be 
determined with an accuracy better than 3\%, independently of the standard IRAC flux calibration (Schuster et al, 2006).

\begin{figure}
\centering
\includegraphics[width=0.85\hsize]{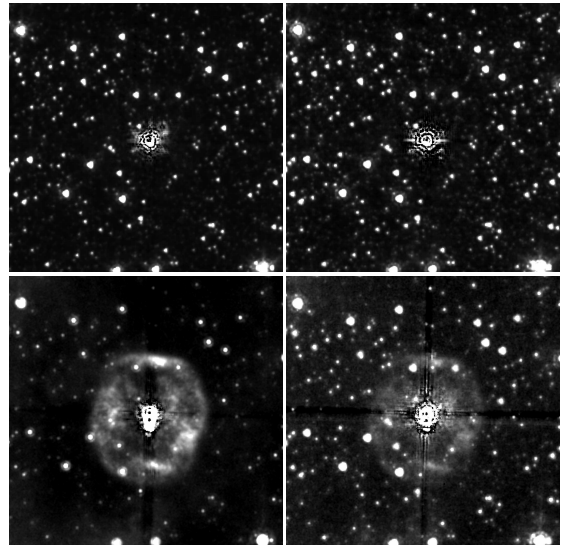}
\caption{IRAC images after PSF subtraction. Clockwise from top left: 3.6, 4.5, 5.8, and 8.0 $\mu$m. Each image is about 200$''$ wide and aligned with north up and east to the left. }
\label{fig:psf_images}
\end{figure}

We thus obtained the photometry of the central source in all of the four IRAC channels. In the 5.8 and 8.0 $\mu$m bands, we clearly detect, beyond the central star, an extended detached shell, brighter at 8.0 than at 5.8 $\mu$m (Figure \ref{fig:psf_images}). The size of the nebula  is about 100$''\times$85$''$. At the assumed distance of 3.5 kpc, this implies a physical size of about $1.70\,$pc$\,\times 1.44\,$pc.
The nebula has an elliptical shape with clumps in both the north-south and (roughly) east-west directions.

\begin{table*}
\centering
\caption{Optical and infrared photometric data for MWC 930 (star + nebula).}
\begin{tabular}{lccccc}
\hline \hline
Optical & B   & V   &  R  &  & \\
S$_\nu$ (Jy)&   0.006   & 0.028   &  0.156  & & \\
   \hline
2MASS & J & H & Ks & & \\     
S$_\nu$ (Jy)&  3.48 & 4.86 & 5.24 & & \\   
  \hline
IRAC  & 3.6 $\mu$m & 4.5 $\mu$m & 5.8 $\mu$m & 8.0 $\mu$m & \\
S$_\nu$ (Jy)&    $8.5 \pm 0.3$   &  $7.5 \pm 0.2$   &  $6.1 \pm 0.2$   &  $5.14 \pm 0.15$  & \\
\hline
WISE  & 3.4 $\mu$m   & 4.6 $\mu$m   & 12 $\mu$m   & 22 $\mu$m   & \\
S$_\nu$ (Jy)&      $ 8.6 \pm 0.9$  & $13 \pm 1$  & $1.6 \pm 0.2$  & $4.3 \pm 0.4$  &  \\
\hline 
AKARI  &9 $\mu$m & 65 $\mu$m & 90 $\mu$m & 140 $\mu$m & 160 $\mu$m \\
S$_\nu$ (Jy)&       $3.25 \pm 0.02$ & $15 \pm 1$ & $12.5 \pm 0.8$ & $8 \pm 1$ & 4.38\tablefootmark{a} \\ 
\hline   
IRAS & 12 $\mu$m & 25 $\mu$m & 60 $\mu$m & 100 $\mu$m & \\ 
S$_\nu$ (Jy)&      $1.8 \pm 0.2$ & $4.0 \pm 0.3$ & $38 \pm 4$ & $35 \pm 3$ & \\
     \hline
\end{tabular}
\tablefoot{\tablefoottext{a}{No error listed in the catalog.}}
\label{tab:data}
\end{table*}

\begin{figure*}
\centering
\includegraphics[width=0.85\hsize]{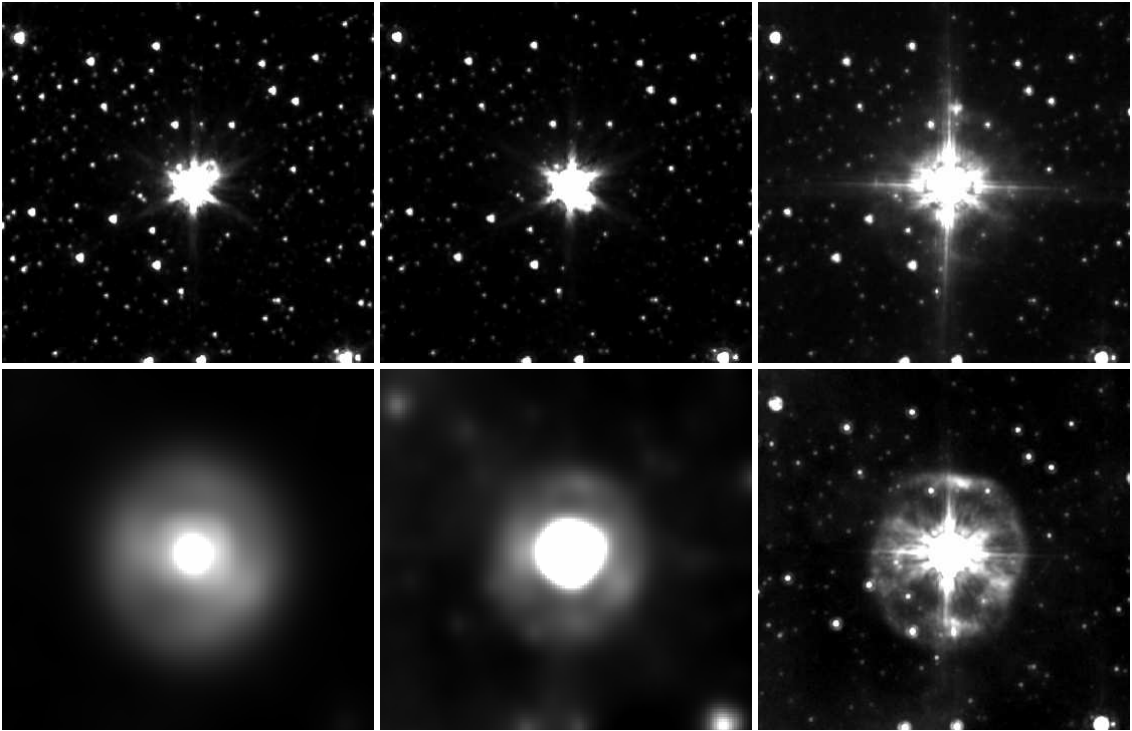}
\caption{IRAC and WISE images of MWC 930. Clockwise from top left: the IRAC images at 3.6, 4.5, 5.8, and 8.0 $\mu$m and then the WISE images at 12 and 22 $\mu$m. Size and orientation as in Figure~\ref{fig:psf_images}.}
\label{fig:images}
\end{figure*}

To better investigate the nature of this detached shell, we searched the web archive of the Wide-field Infrared Survey Explorer \citep[WISE;][]{wise} and retrieved its images of MWC 930 in all bands. While in the first two shorter-wavelength bands, the emission is primarily point-like, at 12 and 22 $\mu$m the detached shell seen with IRAC is detected, as can be seen in Figure~\ref{fig:images}. In these two bands, we performed aperture photometry, since the values listed in the WISE catalog (obtained by PSF fitting) do not include the emission over the whole outer nebula. Our IRAC and WISE photometric data are reported in Table \ref{tab:data}, where the flux densities are referred to the whole emitting source (central star $+$ detached nebula). 

We also performed aperture photometry on the 5.8 and 8.0 $\mu$m IRAC images, after PSF subtraction and could thus calculate the flux density of the detached nebula, as reported in Table~\ref{tab:data2}. 

\begin{table}
\centering \small
\caption{IRAC flux densities of the two components in MWC 930.}
\begin{tabular}{lcccc}
\hline\hline
MWC 930   & 3.6 $\mu$m & 4.5 $\mu$m & 5.8 $\mu$m & 8.0 $\mu$m \\
   & Jy & Jy & Jy  & Jy \\
central source & $8.5\pm0.3$ & $7.5\pm0.2$ & $5.2\pm0.2$ & $4.0\pm0.1$ \\
nebula  & -- & -- & $0.89\pm0.09$ & $1.14\pm0.1$ \\
\hline
\end{tabular}
\label{tab:data2}
\end{table}

\subsection{Spitzer IRS observations}
\begin{figure*}
\centering
\includegraphics[width=0.7\textwidth]{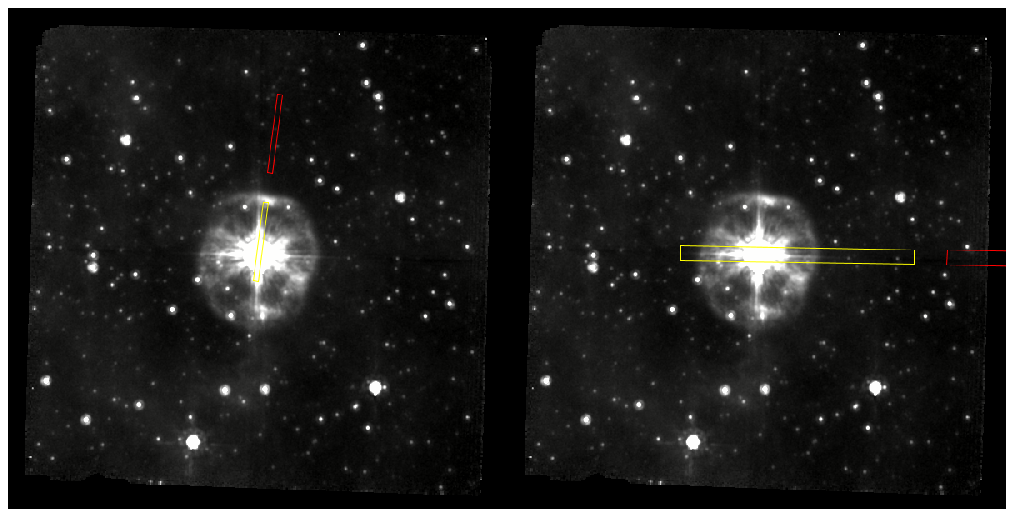}
\caption{Overlay of the IRS slits on the IRAC image at 8 $\mu$m: on the left side the two orders of the SL module and on the right side those of the LL one.}
\label{fig:focalplane}
\end{figure*}

Observations with the IRS were carried out at low spectral re\-so\-lution (R$\sim$60--120). We used both the Short-Low (SL) and the Long-Low (LL) IRS modules. Each module had two orders, which covered the 5.2--7.7 $\mu$m and 7.4--14 $\mu$m ranges at short wavelength  (SL module) and  the 14.0--21.3 $\mu$m and 19.5--38.0 $\mu$m ranges at long wavelength (LL module). The target was observed  with a ramp time of 6 s per cycle and two cycles per module were performed.  We retrieved the IRS BCDs (pipeline version S18.18.0) from the Spitzer archive. As at low resolution the target is observed in two different slit positions, background subtraction was performed by using the observation at one nod position as a background for the other. The spectra were extracted with SMART\footnote{SMART was developed by the IRS Team at Cornell University and is available through the Spitzer Science Center at Caltech.} \citep{higdon} with the optimal extraction algorithm. The IRS slits are overlayed in Figure \ref{fig:focalplane} on the IRAC 8.0 $\mu$m image of MWC 930.

We also extracted the spectra from BCDs where we had subtracted the background in a different way. Each IRS low-resolution module had two orders observing at the same time, but not in the same direction, hence when one was on target, the other would perform a sky observation. We combined these off-target observations together and subtracted them from the corresponding on-target BCDs. The main difference between the two methods of background subtraction is that while the first (nod subtraction) eliminates any contribution to the spectrum from possible circumstellar nebulosity (assuming this is homogeneous within the slit), the second can actually show contributions from a circumstellar nebula. 

For the two different background subtractions, we also extracted the spectra with both the optimal algorithm and over the full slit (appropriate for extended sources). No differences were found between the optimal and full-slit spectra, which indicates that the fraction of flux collected from the detached nebula by the IRS modules is likely negligible. A small difference was instead observed between the spectra extracted from BCDs with different background subtraction.

\begin{figure}[ht]
\centering
\includegraphics[width=0.9\hsize]{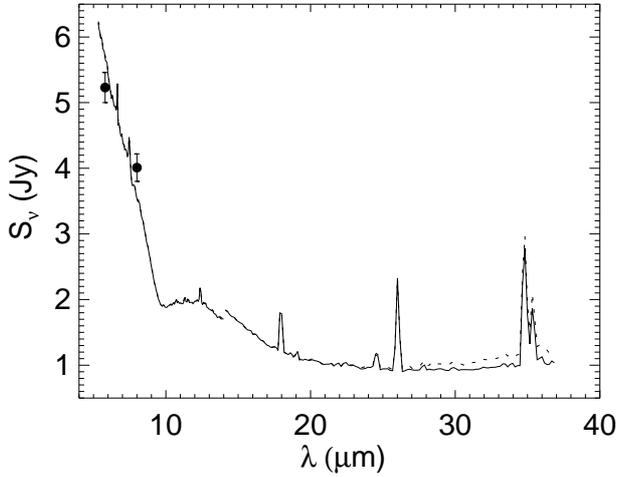}
\caption{IRS spectrum of MWC 930 extracted from nod-subtracted (solid line) and off-subtracted (dotted line) BCDs. IRAC photometric measurements at 5.8 and 8.0 $\mu$m are also displayed.}
\label{compare}
\end{figure}

As can be seen in Figure~\ref{compare}, both the spectrum from the nod-subtracted BCDs and that from the BCDs obtained with the off-target subtraction show an underlying continuum excess at long wavelengths. This appears to be approximately constant with wavelength in the nod-subtracted and increasing with wavelength beyond $\sim$23 $\mu$m in the off-subtracted data. As mentioned above, this means that a small fraction of emission from the circumstellar nebula is caught by the detectors and more efficiently eliminated by the nod subtraction. Below $\sim$23 $\mu$m, the emission within the slit seems dominated by the central source, as shown by the fact that the spectrum is reasonably matched by the IRAC photometric data of the central source alone.

The continuum excess at long wavelengths can be interpreted as due to thermal emission from circumstellar dust. Since only a small fraction of the circumstellar nebula falls within the IRS slit (Figure~\ref{fig:focalplane}), it is not possible to rely on the IRS data to derive the temperature of the emitting dust. 
 
Finally, we extracted the spectrum from the off-subtracted BCDs at positions spaced 2$''$ from one another along the slit. This was done to check whether the nebular emission could be detected moving away from the bright central star. However, our inspection of the spectra did not reveal any detectable nebular radiation.

\section{The IRS spectrum: atomic lines}\label{sec:irsspectrum}
Several lines can be identified in the IRS spectrum of MWC 930. Almost all are due to ionized elements and in particular to \ion{Fe}{ii}. For some lines, the low spectral resolution of our data does not allow us to clearly attribute the feature to one element. In particular, almost all the lines of \ion{He}{ii} and \ion{H}{i} are too close to be distinguished from one another. We notice that in previous studies of LBVs performed with the IRS at higher spectral resolution, HeI lines were detected, but not [\ion{He}{ii}] \citep{hd16,hrcar}.
Furthermore, the \ion{He}{ii} lines would imply a high level of nebular excitation, which is not supported by the detection of highly-excited lines of other species.
A summary of the lines detected is given in Table \ref{tab:features}. 

The inspection of the IRS data, and especially the extraction of the spectrum at different positions along the slit, indicates that the lines do not originate in the extended nebula.  Instead, they are found to arise from the central region. 

\begin{table*}\centering
\begin{tabular}{lcccc}
\hline\hline
Species & Exp. $\lambda_c$ & Obs. $\lambda_c$ &  FWHM  &   Flux \\
        &  $\mu$m          &  $\mu$m  &  $10^{-3}$ $\mu$m & $10^{-13}$ erg cm$^{-2}$ s$^{-1}$\\
        \hline
$[$NiII$]$  & 6.6360   &  6.64 $\pm$ 0.01   &  63 $\pm$ 7     &  30 $\pm$ 11 \\
HeII / HI & 7.45682 &  7.47 $\pm$ 0.01    & 106 $\pm$ 3     & 31 $\pm$ 3 \\
$[$NiII$]$ & 10.6822 & 10.70 $\pm$ 0.01    & 119 $\pm$ 3    &  2.3 $\pm$ 0.2 \\
$[$NiI$]$ & 11.3075 & 11.31 $\pm$ 0.01   & 87 $\pm$ 4  &  2.0 $\pm$ 0.3 \\
HeII / HI  & 12.3669  & 12.37 $\pm$ 0.01   & 118 $\pm$ 4  & 6.8 $\pm$ 0.7 \\
$[$FeII$]$ & 17.9359 &  17.94 $\pm$ 0.01    & 108 $\pm$ 3      & 16 $\pm$ 2 \\
HI      & 18.6152  & 18.61 $\pm$ 0.02  &  179 $\pm$ 6     & 0.67 $\pm$ 0.08  \\
HeII / HI & 19.0498 & 19.04 $\pm$ 0.01   & 114 $\pm$ 11     & 3 $\pm$ 2 \\
$[$FeII$]$ & 24.5193 & 24.55 $\pm$ 0.02    & 296 $\pm$ 22    &  4 $\pm$ 1 \\
$[$FeII$]$ & 25.9883 & 26.02 $\pm$ 0.01    & 254 $\pm$ 2     &  16.7 $\pm$ 0.3 \\
$[$HeII$]$ & 27.8255 & 27.81 $\pm$ 0.03    & 365 $\pm$ 26    &  1.6 $\pm$ 0.4 \\
$[$SiII$]$ & 34.8152 & 34.79 $\pm$ 0.02    & 338 $\pm$ 18     &  17 $\pm$ 3 \\
$[$FeII$]$ & 35.3487 & 35.36 $\pm$ 0.01    & 308 $\pm$ 12     &  7.2 $\pm$ 0.9 \\
\hline
\end{tabular}
\caption{Atomic lines detected in the IRS spectrum.}\label{tab:features}
\end{table*}

Among the lines detected, that of $[$FeII$]$ at 26 $\mu$m and that of $[$SiII$]$ at 34.8 $\mu$m can be compared to the model plots derived by \citet{kaufman1} and \citet{kaufman2} in Photo Dissociation Regions (PDRs).
The average radiation field acting on the PDR around the star can be described by the parameter G, the intensity of the UV radiation field in units of the average interstellar radiation field flux in the far ultraviolet range (from 6 eV to 13.6 eV) \citep{habing}:
\begin{equation}
G=\frac{L \, F_{UV}}{4\pi\,R_i^2\,G_0}
\end{equation}
The parameter $G$ depends on the stellar luminosity $L$, the fraction of energy $F_{UV}$ at
wavelengths shorter than 6 eV ($\sim$2066 $\AA$) in a blackbody spectrum at the same temperature as the star, $R_{i}$ the inner radius from the star to the PDR and $G_0$ the average UV flux in the diffuse ISM \citep[$G_0=1.6 \times 10^{-3}$ erg cm$^{-2}$ s$^{-1}$;][]{draine}. 

Since we have noticed that the lines arise in a region close to the star, not in the outer nebula, we do not have a direct estimate of $R_{i}$. The two IRS slits have widths of $\sim$4$''$ (SL) and $\sim$10$''$ (LL). If we take half of the LL slit width as an upper limit to the inner radius scaled to the assumed distance of 3.5 kpc ($R_i \le 2.62\times10^{17}$ cm), we find $G\sim 6.7\times10^5 G_0$. \citet{gauba} detected a possible nebulosity around the central star and gave 2$''$ as an upper limit to its radius ($R_i \le 1.05\times10^{17}$ cm). Such a value for $R_i$ would give us $G \sim 4.2\times 10^6 G_0$.

\begin{figure*}
\centering
\includegraphics[width=0.45\hsize]{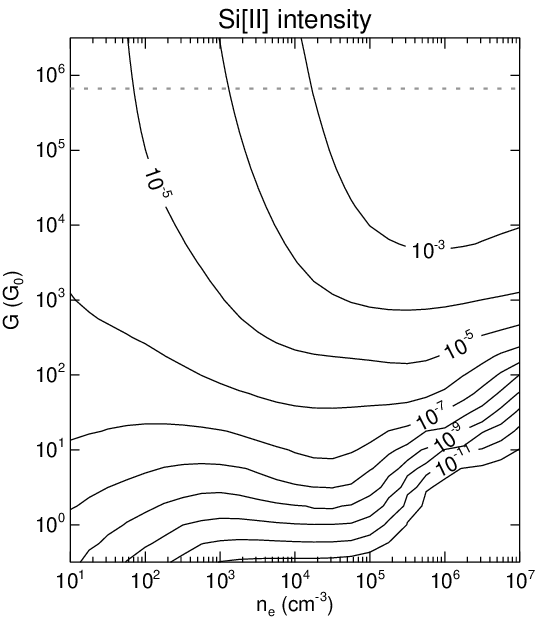}
\includegraphics[width=0.45\hsize]{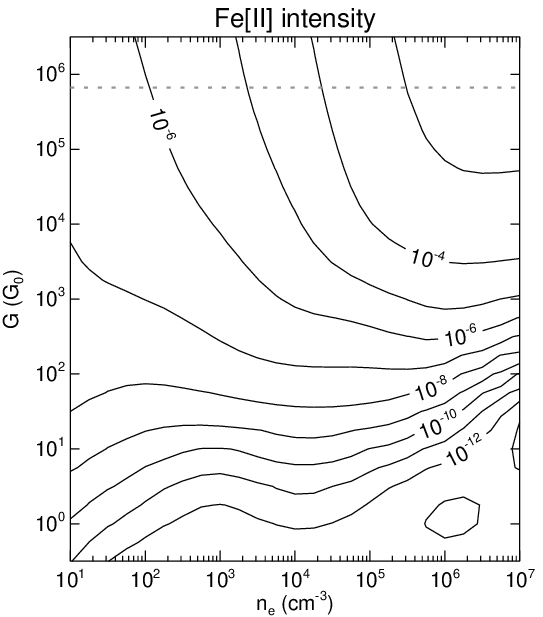}
\caption{Plots of the $[$SiII$]$ and $[$FeII$]$ line intensities in erg~cm$^{-2}$~s$^{-1}$~sr$^{-1}$ as a function of density and local radiation field, from the PDR models by \citet{kaufman2}. The dotted line indicates $G=6.7\times10^5 G_0$.}\label{fig:kaufman}
\end{figure*}

To estimate the density in the PDR, we make use of the plots given by \citet{kaufman2} of the intensity of the $[$SiII$]$ and $[$FeII$]$ lines as a function of $n_e$ and $G$, which we display in Figure~\ref{fig:kaufman}. The model used to obtain the plots had the following elemental abundances: $[$C$]$/$[$H$]=1.4\times10^{-4}$,  $[$O$]$/$[$H$]=3.2\times10^{-4}$,  $[$Si$]$/$[$H$]= 1.7\times10^{-6}$,  $[$S$]$/$[$H$]= 2.8\times10^{-5}$,  $[$Fe$]$/$[$H$]= 
1.7\times10^{-7}$, $[$Mg$]$/$[$H$]= 1.1\times10^{-6}$, $[$PAH$]$/$[$H$]= 6\times10^{-7}$.

If we take 5$''$ as the radius of the emitting region, from the fluxes in Table~\ref{tab:features} we can calculate  the following intensities for the $[$SiII$]$ and $[$FeII$]$ lines, respectively:
$\sim$9.21$\times10^{-4}$ and $\sim$9.05$\times10^{-4}$ erg cm$^{-2}$ s$^{-1}$ sr$^{-1}$ ($5.8\times 10^{-3}$ and $5.6 \times 10^{-3}$ erg cm$^{-2}$ s$^{-1}$ sr$^{-1}$ for an inner radius of 2$''$, respectively).
By combining these values with that of $G$, we can try to estimate the density of the PDR from the model plots. In fact, the $[$SiII$]$ line intensity implies a density of $\sim$2$\times10^4$ cm$^{-3}$, while the $[$FeII$]$ line gives a value of $\sim$3$\times10^5$ cm$^{-3}$, about one order of magnitude larger. Densities of the order of $10^6$ cm$^{-3}$ are reached instead if the inner radius is set to 2$''$.

Two main factors affect the discrepancy between the density values derived from the different lines: the uncertainty in the value of the inner radius of the PDR and the assumption that both lines are excited at the same distance. Assuming 5$''$ is an upper limit for the radius of the PDR, we conclude that the emitting region has a high density at least  of the order of $10^4$ cm$^{-3}$.

\section{The IRS spectrum: solid-state spectral features}\label{sec:solid}
Unlike other LBVs, where Spitzer spectra have led to detect large C-bearing molecules such as polycyclic aromatic hydrocarbons \citep[PAHs; ][]{hd16, hrcar}, MWC 930 does not display signs of these or other carbonaceous components.

One of the clear features detected in the spectrum of MWC 930 is a deep absorption around 10 $\mu$m. A well known broad feature centered around 9.7 $\mu$m is typical of amorphous silicate grains, typically seen in absorption against a bright continuum (like in the Galactic center) or in emission (or self-absorption) in O-rich AGB/post-AGB stars and red supergiants \citep{molster}. 

\begin{figure*}
\centering
\includegraphics[width=0.45\hsize]{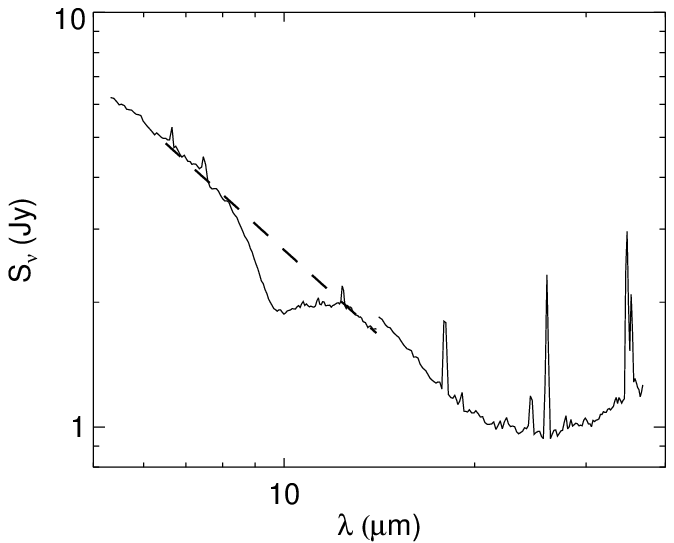} \qquad
\includegraphics[width=0.45\hsize]{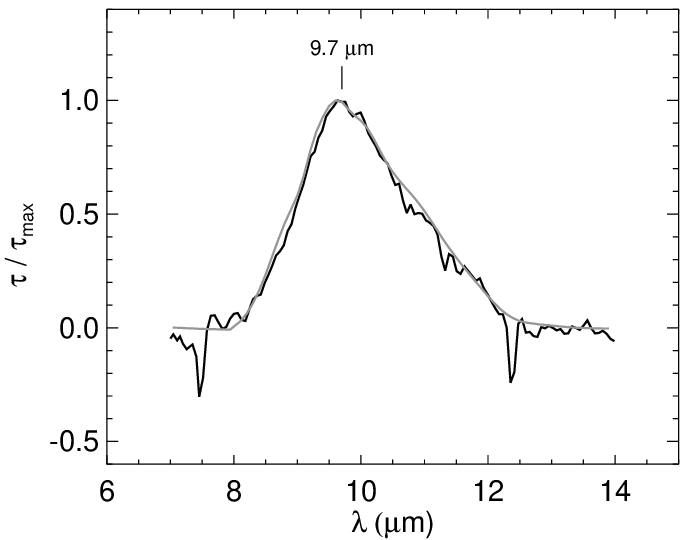}
\caption{\textit{Left:} the silicate feature observed in absorption in the spectrum of MWC 930 with the fitted local continuum represented as a dashed line. \textit{Right:} the optical-depth profile of the same feature (black line) compared to that observed toward the Galactic Center (gray line) by \citet{chiar}.}
\label{fig:silicate}
\end{figure*}

To illustrate the shape of this feature in our target and more easily compare it to similar features in other sources, we have fitted a local continuum (linear in the \textit{log-log} plane), and then derived the optical depth of the feature, considering that $F_{obs}(\lambda)=F_{cont}(\lambda)\, e^{-\tau(\lambda)}$. 
 Figure \ref{fig:silicate} displays the fitted continuum and the resulting optical depth curve. The profile is compared to that given by \citet{chiar} for the silicate feature in the Galactic Center (GC), obtained by the observation of GCS 3 in the Quintuplet Cluster.

In general, the peak wavelength and shape of the feature is due to the exact mixture of silicates (olivines versus pyroxenes, for example) and size of the grains. The presence of crystalline silicates typically makes the shape  change from a peak into a more structured and flatter profile. In MWC 930, we see a relatively smooth profile that  overlaps very well with that seen in absorption toward the GC, therefore we can conclude that it is mostly due to amorphous grains. 

At the same time, the profile in MWC 930 does not differ much from that observed in AGB stars with moderate mass-loss rates ($\sim$10$^{-6}$ $\dot{M}_\odot yr^{-1}$) such as Mira \citep{molster}. In these stars, however, the feature is typically seen in emission from a hot (several 100 K) circumstellar dust layer.

\begin{figure}
\centering
\includegraphics[width=0.85\hsize]{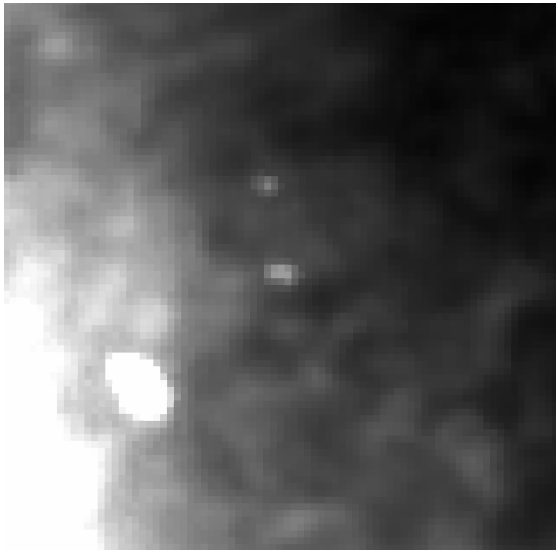}
\caption{Cutout of the IRAS atlas around MWC 930 at 60 $\mu$m ($2^\circ\times2^\circ$). North is up and east to the left.}\label{fig:iras}
\end{figure}

The similarity of the silicate profile in MWC 930 with that of the same feature seen toward the GC, due to interstellar dust, casts some doubt on the feature in our target arising from circumstellar material. 
The Galactic coordinates of MWC 930 are $l \sim$23$^\circ$.65 and $b \sim$2$^\circ$.23; it is therefore relatively close to the Galactic plane. 
In the IRAS Atlas at 60 and 100 $\mu$m, we can see diffuse emission from cold gas and/or dust around our target (Figure~\ref{fig:iras}). We will investigate further the nature of this feature in Section \ref{sec:sed}, when modeling the Spectral Energy Distribution (SED). 

\section{The Spectral Energy Distribution}\label{sec:sed}
To further investigate the properties of the circumstellar environment, we modeled the SED with the 1D code DUSTY \citep{dusty}. For this analysis, we used the photometric data from \citet{miros}, who discussed the variability of the source. From their Figure~1, it can be deduced that both V and K magnitudes vary of 6--7\% over  a $\sim$17-year time. We assumed average values for V and K magnitudes from the data shown by \citet{miros} and derived the other magnitudes accordingly, from the average color indices given in the same work. These data were complemented by our new IRAC and IRS observations as well as by survey measurements from the InfraRed Astronomical Satellite \citep[IRAS]{iras}, WISE and AKARI \citep{akari}, as shown in Figure~\ref{fig:sed}.

The figure displays a shallow energy distribution at near-IR wavelengths. \citet{miros} had already noticed an excess of emission in the near-IR, which they attributed to a compact gaseous nebula around the central star. This can be identified as the same compact nebula that is generating our IRS emission lines and that was detected by \citet{gauba}.

The optical and near-IR observations were dereddened before performing the modeling according to \citet{cardelli}. In doing so, we took into account the values of extinction given by \citet{miros}, who give an interstellar color excess of about 2.5.

The modeling was performed assuming an MRN distribution of the radii of the spherical dust grains \citep{mrn}, with a minimum radius of 0.005 $\mu$m, a ma\-xi\-mum of 0.25 $\mu$m and a power law index of $-3.5$. The optical constants were those for astronomical silicates given by \citet{drainelee}.

\subsection{From UV to mid-IR}
We took as the source of energy a combination of two Planck curves to reproduce the central star and the compact circumstellar gas. Even such a configuration would not allow us to reach a good fit of the infrared data, as excess emission would still be present at IRAC wavelengths. Therefore, we decided to add a third Planck curve and adjusted the relative luminosities and temperatures of the three central sources to best fit the data, keeping the temperature of the central star set to $22\,000$~K.

The outcome of the DUSTY modeling is displayed in Figure~\ref{fig:sed}. The temperatures of the central Planck curves are $22\,000$, $5\,000$ and $1\,100$ K, with contributions to the total luminosity of 95.5\%, 4.2\% and 0.3\%, respectively.  While the extra contribution at $5\,000$ K can be circumstellar gas or even a companion star, as proposed by \citet{miros}, the colder component at $1\,100$ K may be interpreted as due to another gaseous component (an older eruption, for example) or it may point to the existence of hot dust around the star. For a distance of 3.5 kpc, \citet{miros} determines a total luminosity of about $10^{5.6}$ L$_\odot$. We find that, after dereddening, a value of $10^{5.7}$ L$_\odot$ is more adequate to model the data.

\subsection{The far-IR}
\begin{figure}[ht]
\centering
\includegraphics[width=0.9\hsize]{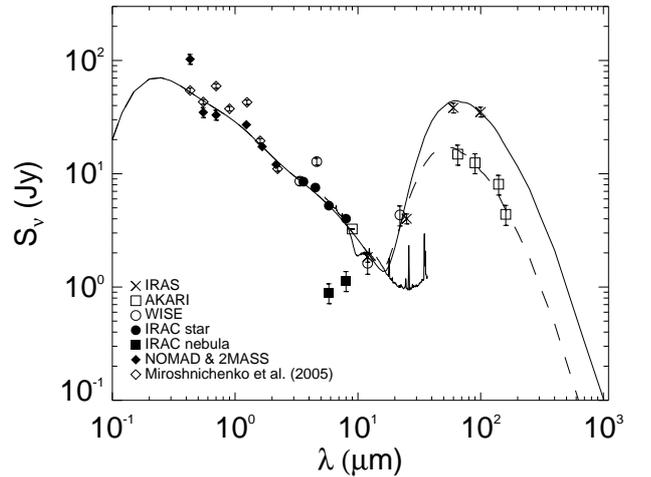}
\caption{SED of MWC 930 with overlayed the DUSTY outputs: a solid line for the model that reproduces the IRAS far-IR data, a dashed line for the model reproducing the AKARI points.}
\label{fig:sed}
\end{figure}

 With the far-IR shape  of the SED shown in Figure~\ref{fig:sed}, it is easy to recognize that any gray body fitting the far-IR points (either the AKARI or IRAS data) will not fit the two IRAC measurements of the nebular emission, falling well below them.
 In fact, the color temperature derived from the two IRAC measurements is $T_c=360 \pm 70$~K, a value much larger than what is expected for the shell generating the far-IR peak. Observations at other wavelengths are desirable to better sample the SED of the nebula detected with IRAC. If the two points are really due to warm dust, it is not clear how this would be kept at high temperature so far away from the central star. 
 
 An alternate possibility is that the IRAC photometric measurements are strongly contaminated by line emission. Around 5.8 and 8.0 $\mu$m, the IRS spectrum is dominated by the radiation from the central source and the examination of the BCDs after background subtraction did not show hints for emission from the detached shell. We are therefore unable to say whether the IRAC fluxes are contaminated by lines.

The AKARI and IRAS observational points are clearly in disagreement. Such a discrepancy may be due to the IRAS photometric measurement being  performed over the whole source, while the AKARI one - with a smaller PSF - is missing extended emission. The FWHM of the Point Spread Functions of the AKARI Far-Infrared Surveyor (FIS) detectors are 37$''$ (at 65 $\mu$m), 39$''$ (90 $\mu$m), 58$''$ (140 $\mu$m) and 61$''$ (160 $\mu$m), which must be compared to a nebular diameter of about 100$''$. It is therefore possible that the PSF-fitting photometry of the AKARI catalog underestimates the emission of the extended nebula. Unfortunately, AKARI images are not yet publicly available to allow us to perform the photometry of our target.  
Furthermore, the values at 65, 140 and 160 $\mu$m are listed in the catalog as \lq\lq low reliability\rq\rq, as the source is not confirmed.

In principle, it might also be true that the IRAS measurements overestimate the emission from the source, because of contamination from nearby material. We showed in Figure~\ref{fig:iras} that the IRAS images display a field rich in diffuse emission.

In the attempt to figure out whether the IRAS or the AKARI data set is more reliable, we modeled the SED twice, once matching the IRAS points and once matching the AKARI ones. 
When fitting the IRAS data points, we obtain a dust temperature of the nebula at its inner radius ($R_{in}$) of 63 K and an opacity in V of $1.1\times10^{-3}$. From the output of DUSTY, we can estimate $R_{in} \sim 3.2 \times 10^{18}$~cm ($\sim$1.04 pc), which at the assumed distance equals $\sim$63$''$. If instead of IRAS, we fit the AKARI far-IR data, then the dust temperature at $R_{in}$ is 75 K, the opacity at 0.55 $\mu$m is $5 \times 10^{-4}$ and  $R_{in}\sim 1.9 \times 10^{18}$~cm ($\sim$0.62 pc), which gives about 38$''$. The dust masses of the nebula also differ in the two cases. Matching the IRAS data results into a dust mass of 0.12 $M_\odot$, while with the AKARI points we obtain 0.02 $M_\odot$.
Neither of the two fits returns an inner radius very close to that observed with IRAC ($\sim$46$''$).

Assuming a standard gas-to-dust mass ratio of 100, the va\-lues of dust mass would  correspond to 12 and 2 M$_\odot$, for the IRAS and AKARI data fit respectively. A substantial amount of mass was then lost when the nebula was ejected, especially if we consider that MWC 930 is not a very luminous LBV. \citet{miros} estimated that its mass on the main sequence was probably around 35 $M_\odot$. Both values of dust mass are quite large for a single eruptive event. It should be considered that the calculation of the mass from the model output depends on the dust opacity at sub-mm wavelengths, therefore if the grain properties are not correct, neither is the derived mass. For example, it is known that in AGB and post-AGB stars the sub-mm dust emissivity goes with $\lambda^{-1}$ rather than  $\lambda^{-1.7}$ as observed in the interstellar medium and assumed in standard sets of optical constants for \textit{astronomical dust} \citep{justtanont, gurtler}. Such a shallower slope may be due to a dif\-ferent chemical composition of the grains or a different size distribution, since larger grains emit more efficiently at large wavelengths. A similar difference for LBVs, where dust grains are relatively \lq\lq fresh\rq\rq, like in AGB stars, would lower our mass estimates of about an order of magnitude.

\subsection{The silicate feature at 9.7 $\mu$m}
\begin{figure*}
\centering
\includegraphics[width=0.9\hsize]{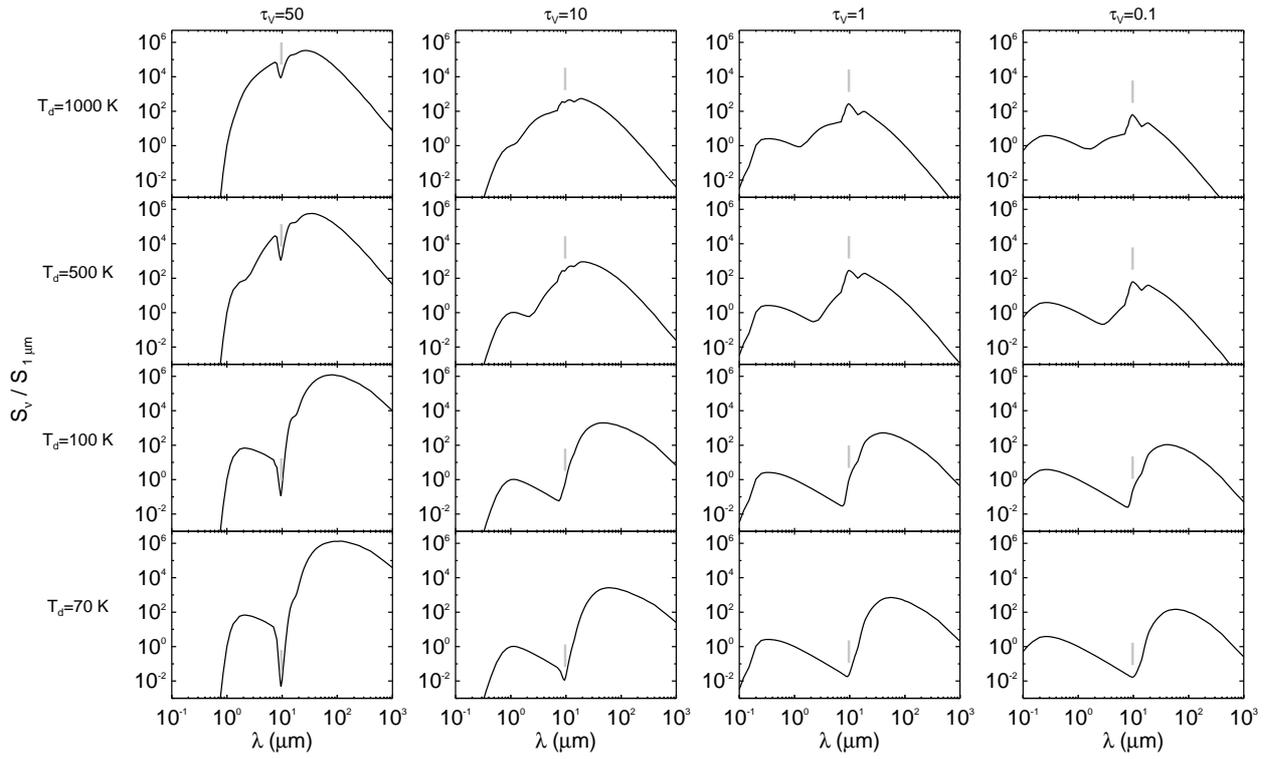}
\caption{DUSTY models obtained at various values of dust opacity and temperature. A gray bar indicates the wavelength of 9.7 $\mu$m in every plot.} \label{fig:models}
\end{figure*}

In Section~\ref{sec:solid}, we noticed that it is not clear whether the silicate absorption feature in the IRS spectrum is of circumstellar or interstellar origin. To figure out which is the case, we will now take advantage of the plots in Figure~\ref{fig:models} obtained with DUSTY for the values of opacity at 0.55 $\mu$m and dust temperature at the inner radius indicated in the figure. The models were calculated taking a Planck curve at $2\times 10^4$ K as the central source, astronomical silicates from \citet{drainelee} to describe the grain properties, and a ratio of outer to inner dust radius of 200.

As we can see, the feature is observed in absorption only at large values of opacity for both hot and cold dust. For small ($< 1$) opacity values, it is only seen in emission, when the dust is hot, or not seen at all, when the dust is cold.

In our source, the feature could be due either to the extended nebula seen with IRAC or to the compact and hot shell at a few arcsec from the star. In either case, the opacity of the dust should be larger than 1, for the feature to be observed in absorption over a hot continuum.

A B-type star, like the central source in MWC 930, emits most of its energy at UV and optical wavelengths, where dust absorption is most efficient. For high opacity values ($\tau_V > 1$), the absorption of the large amount of energy from the central star would determine an intense emission in the far-IR. One would then expect the far-IR peak to be much brighter than the optical/UV peak. This scenario does not match the observational data, where the far-IR peak before dereddening is only slightly brighter than the optical one and after dereddening is actually less bright than the latter. Therefore, the circumstellar opacity in V is unlikely to be larger than 1, which leads us to conclude that the absorption feature is of interstellar origin.

\section{Summary and conclusions}
We have detected a large ($100''\times85''$) circumstellar nebula around the LBV candidate MWC 930  at mid-IR wavelengths, which strengthens the classification of the source as an LBV.

As seen at other wavelengths, the mid-IR spectrum is characterized by lines of $[$FeII$]$ and a continuum displaying infrared excess emission, if compared with that of a star with T$_\mathrm{eff}= 22\,000$ K. Some lines cannot be clearly attributed because of the low spectral resolution of our data, but the examination of the spectrum indicates that the lines are produced in a compact region of a few arcsec around the star, whose existence had been noticed before by \citet{miros} and \citet{gauba}. The comparison of the intensities of the $[$FeII$]$ and $[$SiII$]$ lines with model PDR values implies that this compact shell has a high density of at least $10^4$ cm$^{-3}$.

By 1D modeling of the SED, we conclude that besides the central star, two more components must be taken into account as central sources, one at 5000 K and the other at 1100 K. While the former is certainly gaseous, the latter may actually be hot dust condensing close to the star, which would imply that dust formation is going on in the circumstellar environment. The outer nebula detected with IRAC is much colder (63 K at its inner radius) and seems to contain a substantial amount of the original stellar mass. 


\begin{acknowledgements}
This work is based in part on observations made with the Spitzer Space Telescope, which is operated by the Jet Propulsion Laboratory, California Institute of Technology under a contract with NASA. Support for this work was provided by NASA through an award issued by JPL/Caltech
This publication makes use of data products from the Wide-field Infrared Survey Explorer, which is a joint project of the University of California, Los Angeles, and the Jet Propulsion Laboratory/California Institute of Technology, funded by the National Aeronautics and Space Administration.
This research is based on observations with AKARI, a JAXA project with the participation of ESA.
\end{acknowledgements}


\end{document}